\documentclass[twocolumn,conference]{IEEEtran}
\usepackage{graphicx}

\usepackage{amssymb}
\usepackage{amsmath,amsfonts,amssymb}
\usepackage{verbatim}

\usepackage[bookmarks=false]{}

\IEEEoverridecommandlockouts

\begin{document}
\title{Impact of Mistiming on the Achievable Information Rate of Rake Receivers in DS-UWB Systems}

\author{\IEEEauthorblockN{Chunhua Geng, Yukui Pei, Jiaqi Zhang and Ning Ge}
\IEEEauthorblockA{State Key Laboratory on Microwave and Digital
Communications\\
Tsinghua National Laboratory for Information Science and
Technology\\Department of Electronic Engineering, Tsinghua
University, Beijing 100084, China\\
Email: \{gengch07,zhangjq06\}@mails.tsinghua.edu.cn,
\{peiyk,gening\}@tsinghua.edu.cn }

\thanks{This work is supported by National Nature Science Foundation of China No.
60928001 and 60972019, National Basic Research Program of China
under grant No. 2007CB310608, and the National Science \& Technology
Major Project under grant No. 2009ZX03006-007-02 and
2009ZX03006-009.}}

%\author{\IEEEauthorblockN{Chunhua Geng,
%Yukui Pei, Wujie Wen, Liang Zhu, and Ning Ge}
%\IEEEauthorblockA{State Key Laboratory on Microwave and Digital
%Communications\\
%Tsinghua National Laboratory for Information Science and
%Technology\\
%School of Electronic Engineering,Tsinghua University,Beijing, China}

\maketitle

\begin{abstract}
%\boldmath
In this paper, we investigate the impact of mistiming on the
performance of Rake receivers in direct-sequence ultra-wideband
(DS-UWB) systems from the perspective of the achievable information
rate. A generalized expression for the performance degradation due
to mistiming is derived. Monte Carlo simulations based on this
expression are then conducted, which demonstrate that the
performance loss has little relationship with the target achievable
information rate, but varies significantly with the system
bandwidth and the multipath diversity order, which reflects design
trade-offs among the system timing requirement, the bandwidth and
the implementation complexity. In addition, the performance
degradations of Rake receivers with different multipath component
selection schemes and combining techniques are compared. Among these receivers, the widely used maximal ratio combining (MRC) selective-Rake (S-Rake) suffers the largest performance loss in the
presence of mistiming.

\end{abstract}


% \fi
\IEEEpeerreviewmaketitle

\section{Introduction}
Ultra-wideband (UWB) is promising for wireless high rate and short
range communications \cite{UWBRadio}. Direct-sequence UWB (DS-UWB)
\cite{DS-UWB} has received considerable interest due to its fine
properties of coherent processing of the occupied bandwidth and the
widest contiguous bandwidth \cite{DS advantage}.

To exploit the ample multipath diversity, the Rake reception is
widely employed in DS-UWB systems \cite{SuboptimalReceiver}. Various
types of Rake receivers, like selective Rake (S-Rake) and partial
Rake (P-Rake), are proposed recently \cite{S-Rake}. However, Rake receivers have stringent requirements for timing
accuracy \cite{MRC_EGC}. In practical DS-UWB systems, mistiming due to acquisition
and tracking errors is inevitable, thus its effects on the
performance degradation is worthy of investigation. Several studies have explored this issue
in UWB systems \cite{UWB_throughput}-\cite{BER_mistiming_2}. In
\cite{UWB_throughput}, it is shown that the system throughput
degrades significantly with relatively modest increase in timing
errors over additive white Gaussian noise (AWGN) channels. In
\cite{BER_mistiming_1} and \cite{BER_mistiming_2}, the authors
analyze the bit error rate (BER) degradation induced by mistiming
for both fixed and random channels in UWB systems based on Rake
reception. Compared with throughput and BER, the achievable
information rate, which identifies the maximum mutual information
between the input and output of one communication system, is a more
fundamental measurement for system performance, and is also a subject of 
continuing research in UWB systems \cite{UWB-Capacity}. To the best
of the authors' knowledge, the effect of mistiming on the achievable
information rate of Rake receivers in DS-UWB systems has not been
investigated yet to date.

%transmission rate
%free of errors when efficient error-correcting codes are employed

In this paper, a systematic approach is presented to evaluate the
impact of imperfect timing on Rake receivers in DS-UWB systems from
the perspective of the achievable information rate. The influence of
key system parameters on the performance of various types of Rake
receivers is also investigated. In our analysis, a two-step
procedure is adopted. First, a generalized expression of the system
performance degradation due to timing mismatch is derived. Then
based on this expression, the numerical results are obtained
by averaging over a sufficiently large number of channel
realizations. The major contributions of this paper lie in the
following: (1) As for the widely used maximal ratio combining (MRC) S-Rake receiver, we observe
that the performance degradation has little relationship with the
target information rate, but varies significantly with the
occupied bandwidth and the diversity order, which reflects design
trade-offs among the system timing requirement, the bandwidth and
the implementation complexity. (2) The performance degradation of
various Rake receivers, including MRC S-Rake, MRC P-Rake and equal gain combining (EGC)
P-Rake, are compared. Such comparisons shed light on the robustness
of various multipath component selection schemes and combining
techniques to the variation of system parameters in the presence of
mistiming.

This paper is organized as follows: Section II describes the DS-UWB
system model. In Section III, from the perspective of the achievable
information rate, we derive a generalized expression for the system
performance degradation induced by timing mismatch. In Section IV,
Monte Carlo simulations based on the analytic derivation are
conducted to investigate the influence of some key parameters
on the system performance and compare the performance degradation of
various Rake receivers under mistiming. Section V draws conclusions.

%In this paper, the problem will be addressed by analyzing the
%performance of S-Rake receiver at different synchronization levels.
%
%(TCOM03 EGC/MRC ) In particular, perfect estimates of fading
%parameters and timing information on every branch are required. In
%MRC, both the amplitude and the phase, while in EGC, the phase, of
%all combined multipath components are required at the receiver.
%Hence, it is important to investigate the sensitivity of the
%diversity combining techniques to system imparments.
%
%(timing impact and previous studies) . Previous works study mainly
%the bit error rate (BER) performance with ...
%
%
%(come from VTC01) . A more fundamental type of capacity is the well
%known Shannon capacity in bits/s. Our goal is to develop an
%understanding of the role of various parameters on the Shannon
%capacity of UWB communications with timing errors.

%
%
%
% ... This paper
%quantifies the timing tolerances of UWB transmission with different
%system parameters in various scenarios. The goal is to investigate
%the sensitivity to mistiming by quantifying the capacity loss due to
%synchronization errors. The results here provide meaningful
%implications on the ... under different system settings.

%The results depend on a couple a key channel statistics that can . This general approach to channel capacity applies to
%any channel fading type.
%
%with respect to channel statistics and the Rake parameters
% Given the practical operating conditions and
%system implementation constraints, it helps to guide the design of
%key system parameters.

\section{System Models}
\begin{figure}[!t]
\centering
\includegraphics[width = 9 cm]{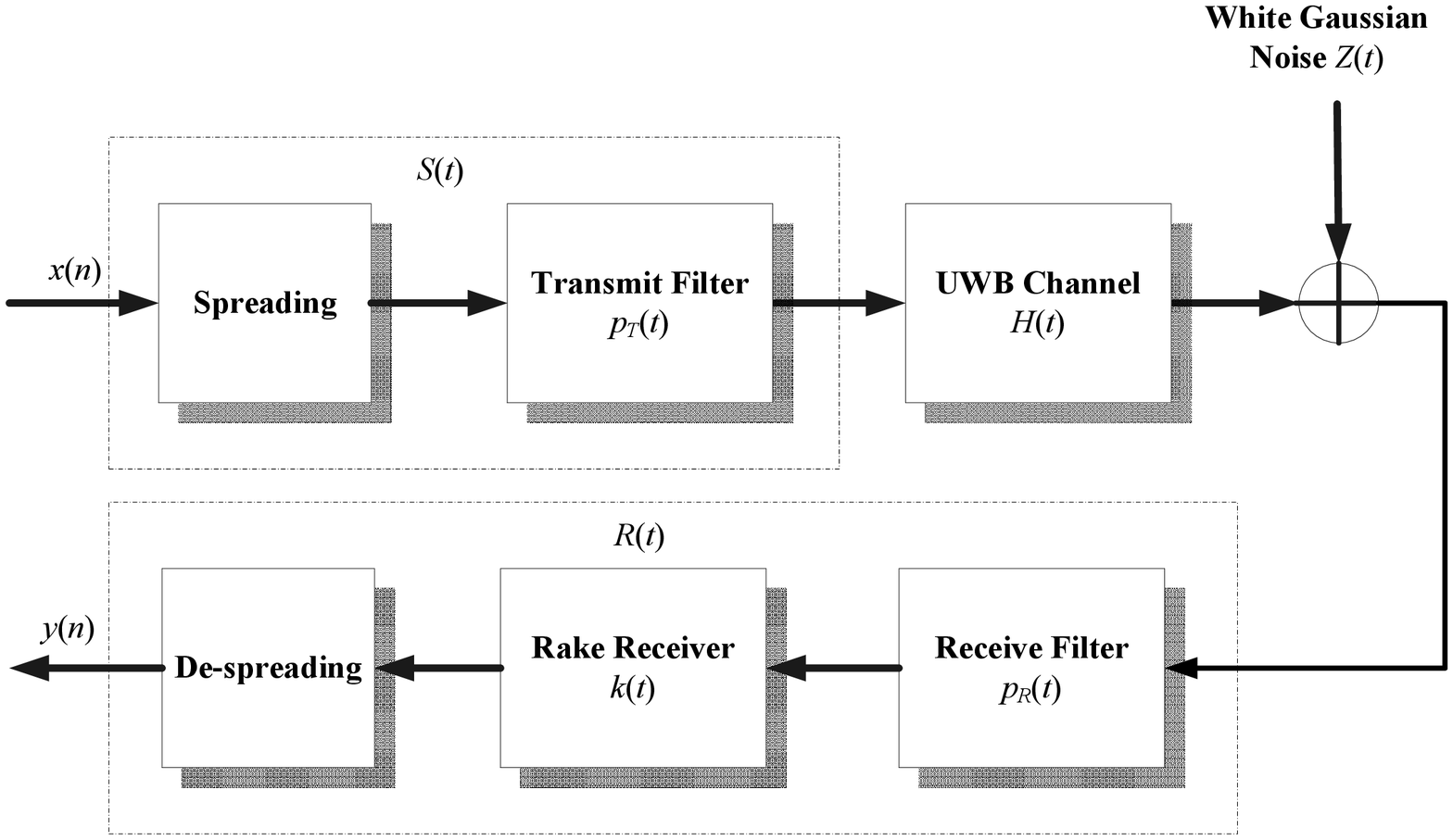}
\caption{Block diagrams for the Rake receiver with pulse shaping
filters in DS-UWB systems} \label{Model2}
\end{figure}

Motivated by current DS-UWB system implementations, we confine our
discussions to binary phase-shift keying (BPSK) modulation. The
equivalent complex-valued system baseband model considered
throughout this paper is shown in Fig.\ref{Model2}.

\subsection{Transmitter Model}

In DS-UWB systems, the random source symbol is spreaded and then modulated with chip pulse
$p_T(t)$. For each symbol, the transmitted waveform is defined as
\begin{equation}
\label{symbol pulse} S(t)=\sum\limits_{n=0}^{N-1} c[n]p_T(t-nT_c)
\end{equation}
where $c[n]$ denotes the $n$-th chip of the spreading code of length
$N$, and $T_c$ is the chip duration.

\subsection{UWB Channel Model}
In this model, the IEEE802.15.3a UWB indoor channel for wireless
personal area networks (WPAN) is considered \cite{Channel}. It
states that the magnitude of channel amplitude better agrees with
the lognormal distribution, corresponding to the shadowing
phenomenon which arises from a more serious fluctuation than
ordinary fading in the impulse response \cite{lognormal}. In
addition, multipath arrivals are grouped into two categories:
cluster arrivals, and ray arrivals within each cluster. The channel
impulse response is defined as:
\begin{equation}
\label{channel}
H(t)=X\sum\limits_{l=0}^{L-1}\sum\limits_{k=0}^{K-1}\alpha_{k,l}\delta(t-T_l-\tau_{k,l})
\end{equation}
where $\delta(t)$ represents the impulse function, $X$ stands for
the log-normal shadowing, $\alpha_{k,l}$ denotes the multipath gain
coefficient, $T_l$ is the delay of $l$-th cluster and $\tau_{k,l}$
is the delay of the $k$-th multipath component relative to the
$l$-th cluster arrival time ($T_l$). By definition, we have
$\tau_{0,l} = 0$ for $l\in\{0,1,...,L-1\}$.

\subsection{Reception Model}
At the receiver, $p_R(t)$ is matched to the impulse response of the
transmit filter $p_T(t)$. In current DS-UWB systems, the raised
cosine filter is commonly employed as the pulse shaping filter,
which is always achieved by implementing root raised cosine filters
as the transmit and receive filters \cite{EQ1}. Therefore,
throughout the rest of this paper, we will consider that the overall
impulse response $p(t)=p_T(t)*p_R(t)$ corresponds to a raised cosine
filter, which means that $p(t)$ can be written as
\begin{equation}\label{4}
    p(t)=p_T(t)*p_R(t)=\frac{\sin(\pi t/T)}{\pi t/T}\frac{\cos(\alpha\pi t/T)}{1-4\alpha^2t^2/T^2}
\end{equation}
where $\alpha$ is the roll-off factor, which represents the excess
bandwidth of the filter and is a real number ranging from 0 to 1.

In this reception model, we assume that the perfect channel
information of the UWB channel is available at the Rake receiver.
The impulse response of the Rake receiver can be written as
\begin{equation}\label{5}
    k(t)=\sum\limits_{j=1}^J w_j\sigma(t-t_j)
\end{equation}
where $J$ stands for the Rake diversity orders i.e. finger numbers,
$w_j$ is the path weights, and $t_j$ denotes the path delay
satisfying $t_j<t_{j+1}$. In P-Rake receiver, the first arrival $J$
multipath components are combined, while the S-Rake receiver selects
out the most $J$ strongest multipath components and then combines
them together. The Rake weights $w_j$ are selected according to
different linear combining techniques. For MRC, $w_j = a^*_j$, while
for EGC, $w_j = sign(\alpha_j)$, where $a_j$ denotes the actual path
amplitude, $(.)^*$ represents complex conjugation, and $sign(.)$ is
the signum function.

Finally, de-spreading is performed to get the symbol-level estimation
of transmitted data $y(n)$. The whole
DS-UWB receiver, including the matched filter $p_R(t)$,
the Rake receiver $k(t)$ and the de-spreading operation, can be expressed as
\begin{equation}
\begin{aligned}\label{recv-impl}
    R(t)= &\sum\limits_{j=1}^J\sum\limits_{n=0}^{N-1} c[n]w_jp_T(t-nT_c-t_j)\\
     = &\sum\limits_{j=1}^Jw_jS(t-t_j)
     \end{aligned}
\end{equation}

Finally, let the impulse response given by $S(t)*H(t)*R(t)$ be
denoted by $g(t)$, and its symbol-sampled version be $g(n)$. Then we
can write $y(n)$ as
\begin{equation}\label{rec_sym}
    y(n)=\sum\limits_{k} x(n-k)g(k) + w(n)
\end{equation}
where $w(n)$ represents the noise component at the Rake output. In (\ref{rec_sym}), $w(n)$
is the symbol-sampled version of $w(t)$, and
\begin{equation}\label{noise}
    w(t)= Z(t)*R(t)
\end{equation}
where $Z(t)$ represents the channel noise which is modeled as AWGN.

\section{Performance Analysis under Mistiming}
In this section, the performance degradation induced by timing
mismatch for Rake receivers in DS-UWB systems is derived in terms of the achievable information rate.

In the DS-UWB system model, when the length of spreading code $N$ is
sufficiently large, the autocorrelation property of spreding code is
ideal. Hence the equivalent channel response between the source
symbols $x(n)$ and the symbol-level received data $y(n)$ can be
simplified to
 \begin{equation}\label{equ-chnnl}
    h(t)=p_T(t)*H(t)*p_R(t)*k(t)
\end{equation}
Its symbol-sampled version is denoted as $h(k)$.

When mistiming is caused by acquisition or tracking errors in the
DS-UWB receiver, the branch delays in the Rake receiver get
inaccurate. Denote this timing mismatch in all branches as
\begin{equation}\label{timing_mismatch}
    \Delta t := t'_j - t_j (\forall j\in\{ 1,2 ... J\})
\end{equation}
where $t_j$ is the actual path delay for path $j$, and $t'_j$ is the
estimated path delay. In this case, the impulse response of the Rake
receiver is given by
\begin{equation}
\begin{aligned}\label{5}
    k'(t)=&\sum\limits_{j=1}^J w_j\sigma(t-t'_j)\\
    =&\sum\limits_{j=1}^J w_j\sigma(t-t_j-\Delta t)\\
    =& k(t-\Delta t)
\end{aligned}
\end{equation}
The corresponding  equivalent channel with timing errors is then
expressed as
 \begin{equation}
 \begin{aligned}\label{equ-chnnl_error}
    f(t)= &  p_T(t)*H(t)*p_R(t)*k(t - \Delta t) \\= &  h(t - \Delta t)
\end{aligned}
\end{equation}
Its symbol-sampled version is written as $f(k)$. This timing
mismatch will result in performance loss in DS-UWB systems.

It is also worthwhile noting that, when the spreading code attains
ideal autocorrelation property and the amplitude modulation schemes,
especially those with bipolar modulation, e.g. BPSK, are employed,
the inter-chip interference (ICI) can be reduced to a negligible
order \cite{AWGN}. Furthermore, to simplify the analysis, we also
assume the excess multipath delay is smaller than several symobl
periods, therefore the effect of inter-symbol interference (ISI) on
the DS-UWB system is also limited. In this case, where the ICI and
ISI are ignorable, the noise component at the Rake output $w(n)$ can
be regarded as AWGN \cite{AWGN}.

In the system model described in section II, the source data are
constrained to be independent and identically distributed (i.i.d).
The system achievable information rate, which corresponds to the
maximum mutual information between the input $x(n)$ and the output
$y(n)$, is given by \cite{IR} \footnote{When the white Gaussian
noise at the Rake output is not valid, i.e. in the case of colored
Gaussian noise, the calculation of achievable information rate can
be performed by using the method of water pouring
\cite{water-pouring}. }
\begin{equation}
\label{information_rate}
    C=\frac{1}{4\pi}\int\limits_{-\pi}^{\pi}\log_2[1+2\frac{E_s}{N_0}|
    H(e^{j\theta})|^2]d\theta
\end{equation}
where $E_s$ is the symbol energy, and $H(e^{j\theta})$ is the
Fourier transform of the equivalent channel impulse response. In the
scenario of perfect synchronization, $H(e^{j\theta})$ stands for the
Fourier transform of $h(k)$; in the scenario with timing errors,
$H(e^{j\theta})$ represents the Fourier transform of $f(k)$.

From (\ref{information_rate}), it is obvious that the achievable
information rate $C$ is derived from $E_s/N_0$ considerations, and a
certain $E_s/N_0$ is required to achieve a specified $C$. Let $R$ be
the target achievable information rate. In the perfect
synchronization scenario, we have
\begin{equation}\label{R_perfect}
C|_{H(e^{j\theta})=F[h(k)]}=R
\end{equation}
where $F\{ \}$ represents the Fourier transform operator. Assume the
needed $E_s/N_0$ in dBs at this point is $SNR_h$.

AS for the scenario with timing errors, let $SNR_f$ be the $E_s/N_0$
in dBs at which
\begin{equation}\label{R_timing_error}
C|_{H(e^{j\theta})=F[f(k)]}=R
\end{equation}

Finally, the performance degradation $L$ induced by timing mismatch is
obtained as
\begin{equation}\label{gap}
    L = SNR_f - SNR_h
\end{equation}

\section{Numerical Results and Discussions}
In this section, Monte Carlo simulations based on the generalized
expressions (\ref{information_rate}) and (\ref{gap}) are conducted
to evaluate the effect of various system configurations on the
performance degradation $L$ induced by timing mismatch in DS-UWB
systems. In the following simulations, in order to keep the
simulation complexity on a reasonable level, the timing mismatch
$\Delta t$ on Rake diversity branches is set as $(0, 0.1, 0.2, ...
0.9) \times T_s$, where $T_s$ denotes the duration of one symbol. In
addition, the numerical results are averaged over the best 900 out
of 1000 IEEE 802.15.3a CM1 channel realizations, following the
recommended instructions in \cite{Channel} that the worst 10\% channels are ignored in the simulation. The rest of this section consists of two parts: In the first
part, the influence of timing mismatch on the performance of the
widely used MRC S-Rake receiver is investigated in details under
different system parameters; In the second part, we compare the
performance loss of various Rake receivers, including MRC S-Rake,
MRC P-Rake and EGC P-Rake, in the presence of mistiming. 

\begin{figure}[!t]
\centering
\includegraphics[width = 9 cm]{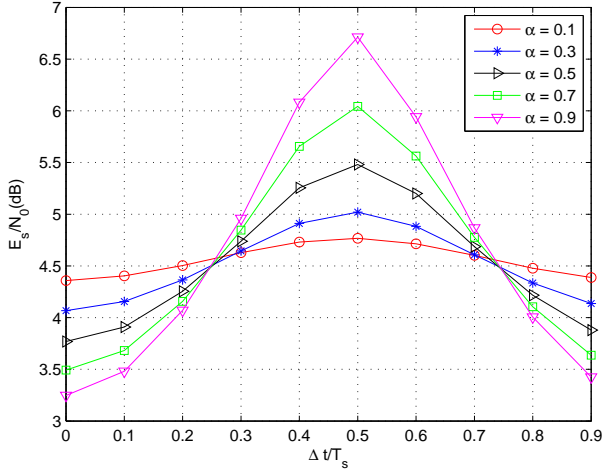}
\caption{The needed $E_s/N_0$ (dB) for MRC S-Rake receivers in
DS-UWB systems with different roll-off factors $\alpha$ when $J = 8$
and $R = 0.3$} \label{MRC S-Rake roll}
\end{figure}

{\it \textbf{The Influence of Timing Mismatch on MRC S-Rake
Receivers}}: The needed $E_s/N_0$ curves for MRC S-Rake receivers in
DS-UWB systems with different roll-off factors $\alpha$ are plotted
in Fig.\ref{MRC S-Rake roll}. It is observed that if no timing
mismatch exists, as $\alpha$ increases, the needed $E_s/N_0$ to
achieve the target information Rate $R$ gets smaller. However, the
receiver with larger $\alpha$ is more sensitive to the timing
mismatch $\Delta t$. As seen in this figure, when mistiming is
small, the receiver with larger $\alpha$ needs less $E_s/N_0$ to
obtain the target $R$; however, they requires more $E_s/N_0$ as
mistiming aggravates.

\begin{figure}[!t]
\centering
\includegraphics[width = 9 cm]{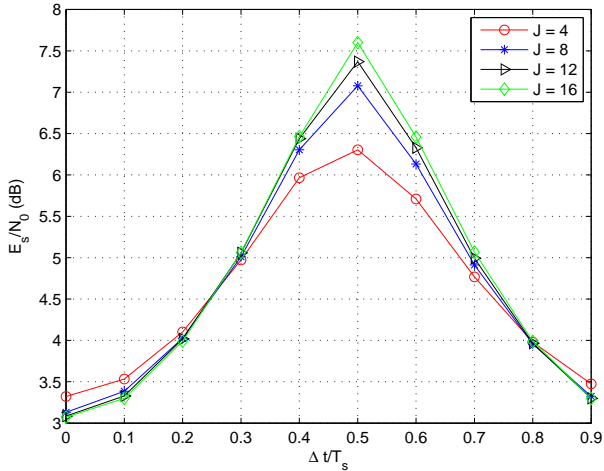}
\caption{The needed $E_s/N_0$ (dB) for MRC S-Rake receivers with
different diversity orders $J$ when $\alpha = 1.0$ and $R = 0.3$}
\label{MRC S-Rake finger}
\end{figure}

Fig.\ref{MRC S-Rake finger} depicts the $E_s/N_0$ needed for MRC
S-Rake receivers with different diversity orders to achieve $R =
0.3$ when $\alpha$ equals to 1.0. As one can expect, when no timing errors
exhibit, the MRC S-Rake receiver with more diversity branches needs
less $E_s/N_0$ to achieve the target $R$. It is further seen that
the sensitivity to timing mismatch increases with increasing $J$ in
the MRC S-Rake receiver. Hence there also exists a design trade-off
between the robustness to mistiming and the implementation
complexity.

The needed $E_s/N_0$ for MRC S-Rake receivers to achieve various
target information rates $R$ is illustrated in Fig.\ref{MRC S-Rake
rate} when $J$ is 8 and $\alpha$ equals to 0.3. It shows that as $R$
increases, the needed $E_s/N_0$ increases obviously. However, the
performance loss induced by timing mismatch rarely varies with
the increase of the target achievable information rate in MRC S-Rake
receivers.

\begin{figure}[!t]
\centering
\includegraphics[width = 9 cm]{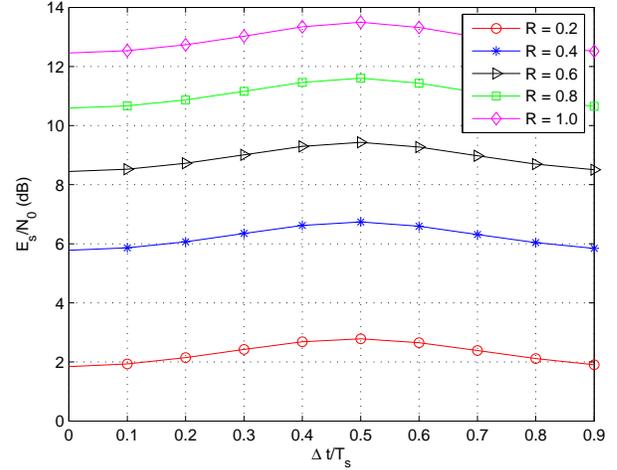}
\caption{The needed $E_s/N_0$ (dB) for MRC S-Rake receivers with
different target achievable information rates $R$ when $\alpha =
0.3$ and $J = 8$} \label{MRC S-Rake rate}
\end{figure}

{\it \textbf{The Comparison of Various Rake Receivers under
Mistiming}}: In this part, the mistiming is
assumed to follow uniform distribution \cite{time-uniform-distr}. We
consider two cases: one is the worst case with the maximum
performance degradation; the other is the average case,
which represents the average degradation over the set of timing
mismatch $\Delta t$.

\begin{figure}[!t]
\centering
\includegraphics[width = 9 cm]{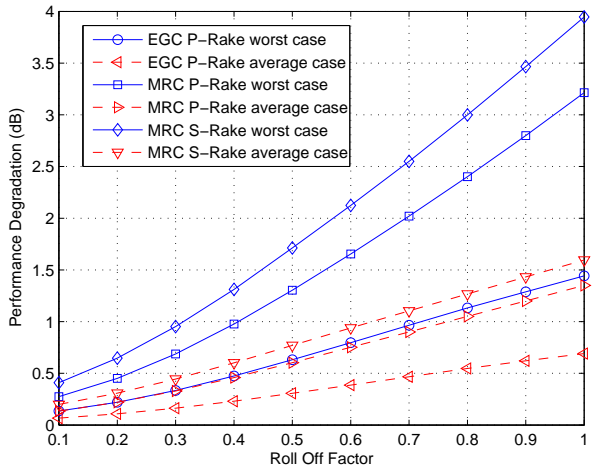}
\caption{Performance degradation $L$ (dB) as a function of the
roll-off factor $\alpha$ for various Rake receivers when $J = 8$ and
$R = 0.3$ } \label{total-roll}
\end{figure}

\begin{figure}[!t]
\centering
\includegraphics[width = 9 cm]{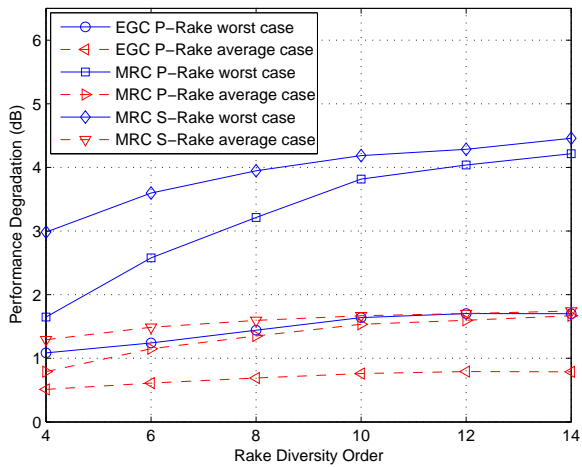}
\caption{Performance degradation $L$ (dB) as a function of the
diversity order $J$ in various Rake receivers when $\alpha = 1.0$
and $R = 0.3$} \label{total-finger}
\end{figure}

\begin{figure}[!t]
\centering
\includegraphics[width = 9 cm]{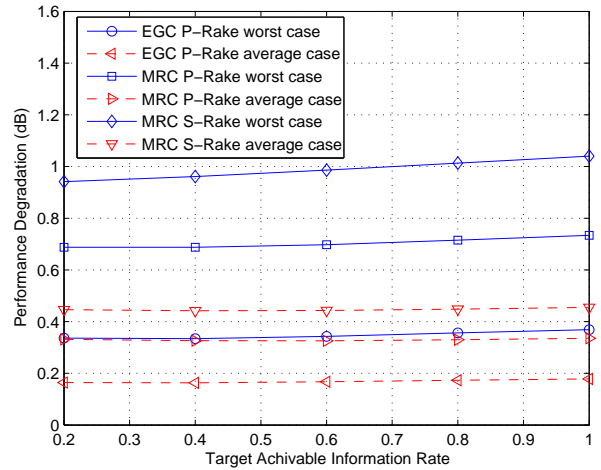}
\caption{Performance degradation $L$ (dB) as a function of the target
achievable information rate $R$ for various Rake receivers when
$\alpha = 0.3$ and $J = 8$} \label{total-rate}
\end{figure}

Fig.\ref{total-roll} - Fig.\ref{total-rate} demonstrate
the performance degradation of three types of Rake receivers,
including MRC S-Rake, MRC P-Rake and EGC P-Rake, with respect to the
variation of the excess bandwidth, the diversity order and the
target achievable information rate under timing mismatch respectively. Fig.\ref{total-roll} shows that with roll-off factor
increasing, all types of Rake receivers obtain worse performance due
to timing errors. Among all the receivers, MRC S-Rake is most
sensitive to the increase of excess bandwidth and the EGC P-Rake is
least sensitive. From Fig.\ref{total-finger}, it is observed that as the
diversity order increases, the performance degradation of all the three
kinds of Rake receivers gets larger. Fig.\ref{total-finger} also
shows that when MRC is employed, P-Rake is more sensitive to the
change of Rake finger numbers than S-Rake, and in P-Rake receiver
the EGC technique is more robust compared with MRC.
Fig.\ref{total-rate} demonstrates that the target information rate
rarely impacts the performance loss of all the Rake receivers under
timing mismatch, and the MRC S-Rake suffers the largest performance
loss compared with other Rake receivers.

\section{Conclusion}

The effect of imperfect timing has been evaluated for the Rake
reception in DS-UWB systems from the perspective of the achievable
information rate. A generalized expression for the system performance degradation is derived,
then corresponding simulations are conducted to investigate the
effect of timing mismatch on the widely used MRC S-Rake receiver
with respect to the excess bandwidth induced by the roll-off factor
in RRC filters, the multipath diversity order and the target
information rate. Simulation results illustrate that the performance
loss has little relationship with the target information rate, but
varies significantly with the system bandwidth and the
diversity order, which further demonstrates that there exist
fundamental trade-offs among the system timing requirement, the
occupied bandwidth and the implementation complexity of DS-UWB systems. In
addition, the performance degradation of various types of Rake
receivers, including MRC S-Rake, MRC P-Rake and EGC P-Rake, is
compared, and the sensitivity of different multipath component
selection schemes and diversity combining techniques to the
variation of system parameters are obtained. The numerical results
also show that among the three types of Rake receivers, the MRC S-Rake
suffers the most performance degradation in the presence of
mistiming.

%(be conducted further for the efficient design of the DS-UWB
%systems) (the tradeoff between ... could facilitate the design of
%)(The performance improvement through the addition of ... is only
%marginal)(The main contribution of this paper lies in the
%following.)

%Compared with EGC, MRC technique is more sensitive to the variation
%of excess bandwidth caused by roll-off factor $\alpha$. In addition,
%the diversity order $J$ in S-Rake receiver exerts stronger influence
%on the performance loss of MRC technique than that of EGC technique.

%\section*{Acknowledgment}




%\bibliographystyle{IEEEtran}

\begin{thebibliography}{1}
\bibitem{UWBRadio}
S. Roy, J. R. Foerster, V. S. Somayazulu, and D. G. Leeper,
"Ultrawideband radio design: the promise of high-speed, short-range
wireless connectivity," {\it Proceedings of the IEEE}, vol.92, no.2,
pp. 295-311, Feb. 2004.


\bibitem{DS-UWB}
R. Fisher, R. Kohno, M. McLaughlin, et al. DS-UWB physical layer
submission to IEEE 802.15 task group 3a (Doc. Number
P802.15-04/0137r4), IEEE P802.15, 2005.

\bibitem{DS advantage}
P. Runkle, J. McCorkle, T. Miller, and M. Welborn, "DS-CDMA: the
modulation technology of choice for UWB communications", IEEE
Conference on Ultra Wideband System and Technologies ({\it UWBST}),
pp.364-368, 2003.

\bibitem{SuboptimalReceiver}
J. D. Choi, and W. E. Stark, "Performance of ultra-wideband
communications with suboptimal receivers in multipath channels,"
{\it IEEE Journal on Selected Areas in Communications}, vol.20,
no.9, pp. 1754-1766, Dec. 2002.

\bibitem{S-Rake}
D. Cassioli, M. Z. Win, F. Vatalaro, and A. F. Molisch, "Low
Complexity Rake Receivers in Ultra-Wideband Channels," {\it IEEE
Transactions on Wireless Communications}, vol.6, no.4, pp.1265-1275,
Apr. 2007.

\bibitem{MRC_EGC}
Y. Yin, J. P. Fonseka, and I. Korn, "Sensitivity to timing errors in
EGC and MRC techniques", {\it IEEE Transactions on Communications},
vol.51, no.4, pp.530-534, Apr. 2003.

\bibitem{UWB_throughput}
W. M. Lovelace, and J. K. Townsend, "The effects of timing jitter
and tracking on the performance of impluse radio", {\it IEEE Journal
on Selected Areas in Communications}, vol.20, no.12, pp.1646-1651,
Dec. 2002.

\bibitem{BER_mistiming_1}
Z. Tian, and G. B. Giannakis, "BER sensitivity to mistiming in
ultra-wideband impluse radios - part I: nonrandom channels", {\it
IEEE Transactions on Signal Processing}, vol.53, no.4, pp.1550-1560,
Apr. 2005.

\bibitem{BER_mistiming_2}
Z. Tian, and G. B. Giannakis, "BER sensitivity to mistiming in
ultra-wideband impluse radios - part II: fading channels", {\it IEEE
Transactions on Signal Processing}, vol.53, no.5, pp.1897-1907, May
2005.

\bibitem{UWB-Capacity}
N. Guney, H. Deli, and F. Alagoz, "Achievable information rates of
PPM impulse radio for UWB channels and rake reception" , { \it IEEE
Transactions on Communications}, vol.58, no.5, pp.1524-1535, May
2010.

%\bibitem{Proakis}
%J. G. Proakis, ed., {\it Digital Communications}. New York, NY,
%McGraw-Hill, Inc., 4th Ed., 2001.

\bibitem{Channel}
Channel-Modeling-Subcommittee-Report-Final, IEEE P802.15. Dec. 2002.

\bibitem{lognormal}
R. C. French, "The effect of fading and shadowing on channel reuse
in mobile radio", {\it IEEE Transactions on Vehicular Technology},
vol.28, no.3, pp.171-181, Aug. 1979.

\bibitem{EQ1}
A. Parihar, L. Lampe, R. Schober, and C. Leung, "Equalization for
DS-UWB Systems¡ªPart I: BPSK Modulation," {\it IEEE Transactions on
Communications}, vol.55, no.6, pp.1164-1173, Jun. 2007.




\bibitem{AWGN}
J. Zhang, R. A. Kennedy, and T. D. Abhayapala, "Performance of Rake
reception for ultra-wideband signals in a lognormal-fading channel",
{\it Proc. IWUWBS}, pp. 5-9, 2003.




%\bibitem{Symbol-Sampled}
%R. D. Souza, and J. Garcia-Frias, "Performance of symbol-sampled
%receivers over unknown continuous-time Rayleigh channels," {\it IEEE
%Transactions on Wireless Communications}, vol.4, no.5, pp.
%2020-2026, Sept.2005.

%\bibitem{SS-VTC}
%R. D. Souza, and J. Garcia-Frias, "On the Performance of
%Symbol-Sampled Receivers Over Unknown Continuous-Time Channels,"
%2006 IEEE 64th Vehicular Technology Conference (VTC-2006 Fall),
%pp.1-4, Sept. 2006.

\bibitem{IR}
W. Hirt, and J. L. Massey, "Capacity of the discrete-time Gaussian
channel with intersymbol interference," {\it IEEE Transactions on
Information Theory}, vol.34, no.3, pp.380-388, May 1988.

%\bibitem{Forney}
%G. Forney Jr., "Maximum-likelihood sequence estimation of digital
%sequences in the presence of intersymbol interference," {\it IEEE
%Transactions on Information Theory}, vol.18, no.3, pp. 363- 378, May
%1972.

\bibitem{water-pouring}
W. Xiang, and S. S. Pietrobon. "On the capactity abd normalization
of ISI channels", {\it IEEE Transactions on Information Theory},
vol.49, no.9, pp.2263-2268, Sep. 2003.

\bibitem{time-uniform-distr}
M. O. Sunay, and P. J. Mclane, "Diversity combining for DS CDMA
systems with synchronization errors", IEEE International Conference
on Communications (ICC), pp.83-89, 1996.






\end{thebibliography}

\end{document}